\documentclass{aastex61}

\accepted{\today}
\submitjournal{ApJ}

\shorttitle{Radiative Transfer in Translucent Clouds}
\shortauthors{Biganzoli, Potenza \& Robberto}
\usepackage{amsmath,amstext}
\usepackage{newtxmath} %use times font for math

\begin{document}
\title{Radiative Transfer in a  Translucent Cloud Illuminated by an Extended Background Source}

\correspondingauthor{Massimo Robberto}
\email{robberto@stsci.edu}

\author{Davide Biganzoli}
\affiliation{Universit\`a degli Studi dell'Insubria \\
Dept. of Science and High Technology\\ 
Via Valleggio, 11, I-22100 Como, Italy}
\affiliation{Universit\'a degli Studi di Milano\\
Dept. of Physics\\ 
Via Celoria 16, 20133 Milano, Italy}

\author{Marco A. C. Potenza}
\affiliation{Universit\'a degli Studi di Milano\\
Dept. of Physics\\ 
Via Celoria 16, 20133 Milano, Italy}

\author[0000-0002-9573-3199]{Massimo Robberto}
\affiliation{Space Telescope Science Institute\\ 
Baltimore, MD 21218}
\affiliation{Center for Astrophysical Studies\\ 
Johns Hopkins University\\
Baltimore, MD 21218}

\begin{abstract}
We discuss the radiative transfer theory for translucent clouds illuminated by an extended background source. First we derive a rigorous solution based on the assumption that multiple scattering produce an isotropic flux. Then we derive a more manageable analytic approximation showing that it nicely matches the results of the rigorous approach. To validate our model, we compare our predictions with accurate laboratory measurements for various types of well characterized grains, including purely dielectric and strongly absorbing materials representative of astronomical icy and metallic grains, respectively, finding excellent agreement without the need of adding free parameters.
We use our model to explore the behavior of an astrophysical cloud illuminated by a diffuse source with dust grains having parameters typical of the classic ISM grains of \cite{Draine} and protoplanetary disks, with an application to the dark silhouette disk 114-426 in Orion Nebula. We find that the scattering term modifies the transmitted radiation, both in terms of intensity (extinction) and shape (reddening) of the spectral distribution. In particular, for small optical thickness our results show that scattering makes reddening almost negligible at visible wavelengths. Once the optical thickness increases enough and the probability of scattering events become close to or larger than 1, reddening becomes present but appreciably modified with respect to the standard expression for line-of-sight absorption. Moreover, variations of the grain refractive index, in particular the amount of absorption, also play an important role changing the shape of the spectral transmission curve, with dielectric grain showing the minimum amount of reddening.
\end{abstract}


\section{Introduction}
The increase in sensitivity, spatial resolution and areal coverage of infrared instrumentation
has enabled in recent years new detailed studies of translucent clouds, defined as clumps of 
interstellar material with optical depth in the range 1 to 5 at some observable wavelength \citep{vanDishoeck+Black}. The intermediate opacity  of 
these systems produces measurable attenuation and reddening of background sources, without completely precluding their observation. 
If the background source density is high enough, or if the brightness of a diffuse background is uniform enough,  it is possible to reconstruct the density profile of the cloud for an assumed reddening law. Density maps of translucent clouds have been obtained for a variety of systems, like e.g. the  Bok globule Barnard 68 \citep{Alves},  the variety of translucent clouds detected by Spitzer \citep{Spitzer} and Herschel \citep{Herschel}, the giant dark silhouette disk 114-426 in the Orion Nebula Cluster \citep{Shuping et al., Miotello et al}. Comparing density maps taken at different wavelengths allows deriving the abundances of indidual species (molecules, ices, dust grains) and their growth \citep{HerschelPRIMAS}.

The analysis of the these systems is normally carried out applying the familiar expressions for the interstellar reddening toward a point source, i.e. the exponential decay law known as Lambert-Beer-Bouguer (LBB) law. However, when the source of background illumination is extended and diffuse, the analysis of the results may require some special attention from the point of view of radiative transfer. 
Let us take as an illustration the case of the giant dark silhouette disk 114-426 in the Orion Nebula.  This source is close enough to the luminous backdrop of the Orion Nebula that an appropriate treatment of the radiative transfer through the outer parts of the disk must take into account the fact that each particle is illuminated by an extended, diffuse source. Extended  illumination means that radiation propagating in every direction can be scattered towards the observer, thus partially compensating for the direct extinction along the line of sight. In cases like this, a one-dimensional treatment where there is only absorption along the line of sight may not be entirely adequate. %Moreover, this effect introduces a fundamental difference between the behavior of pure dielectric (i.e. perfectly reflecting) vs. absorbing scatterers.
\\ In this paper we develop a general, simplified treatment of the scattering by a cloud illuminated by extended, diffuse radiation. In Section 2 we present an analytic solutions of the Radiative Transfer Equation (RTE) that can be 
used in a wide range of astrophysical conditions.
%useful in many cases of astrophysical interest. 
In Section 3 we derive a simplified analytic expression to easily estimate the radiative transfer in the most interesting cases. In Section 4 we validate our models using an experimental apparatus specifically designed to reproduce  an extended, diffuse, white light source, illuminating a sample cell containing water suspensions of well known scatterers. In particular, we report the results obtained with calibrated, monodisperse polystyrene spheres at different concentrations and with polydisperse, nonspherical particles, representative of purely dielectric and strongly absorbing materials, respectively. Finally, in Section 5 we explore how the scattering terms affect the spectral intensity transmitted by a translucent cloud with typical astronomical grains, using grain mixtures appropriate for the ISM and young circumstellar disks, with an application to the 114-426 disk. The Appendix details the methods adopted to achieve absolute, independent characterization of our dust samples, a fundamental step to %work without any free parameter to assess the validity of this study. 
assess the validity of our experimental results without tuning any free parameter.

%Data are compared to the transmission expected on the basis of just the Lambert-Beer-Bouguer (LBB) law. 
%In the second part we corroborate our model presenting transmission measurements performed with an apparatus specifically designed to r. By changing the size, composition, shape of the scatterers, as well as the illumination conditions, it is possible to 
%accurately test the model. A sensitivity analysis based on independent measurements of phase functions and spectral extinction is finally presented  to assess the importance of each contribution to the final result.

\section{Radiative transfer inside a cloud}
The general problem of describing the radiative transfer inside a cloud illuminated by an extended source emitting diffuse radiation has been solved analytically in several ways. Analytic solutions have been obtained assuming Rayleigh scattering \citep{Gilbert}, or making use of iterated integrals to compute Chandrasekhar's functions $S$ and $T$ \citep{chandra} which give the transmitted and reflected radiation  \citep[e.g.][]{Tanaka7,Tanaka10}. Similar problems have been addressed by Chandrasekhar himself, and solved later in several ways, as shown for example in \cite{Liuo, Chalhoub, Barman}. The \textit{discrete ordinates} method, in particular,  gives a solution by expanding the phase function in Legendre polynomials. This approach is extremely powerful, but the analytic expression for the solution is very complex and difficult to handle. Other approaches make use of numerical solutions, e.g.  \cite{Siewert, Herman, Jablonski}. 
%\citet{aur82}. A clear formulation of the problem and an  overview of the approaches adopted so far is given in the recent review by \cite{bailey}, addressing the atmospheres of extrasolar planets and brown dwarfs and the issue of the corresponding radiative tranfer. 

Here we introduce a simplified analytic model that can be conveniently used to  describe light scattering from a cloud of randomly distributed particles illuminated by an extended source. Our goal is to obtain the radiation flux emerging in a given direction, to be detected at a very large distance from the cloud. We give a solution in terms of the typical quantities exploited in radiative transfer models: 1) the phase function $p(\theta)$, 2) the optical thickness of the system, $\tau$, i.e. the length over which intensity is attenuated by a factor $1/e$; and the single scattering albedo, $\omega=\frac{1}{4\pi}\int p(\theta)\sin{\theta}d\theta d\varphi$, the average of the phase function over the sphere. The way these parameters can be determined will be discussed below. 

Let's start by recalling the general formulation of the RTE, in the form used to describe a monodimensional cloud of particles extended along the $z$ axis, that also represents the line of sight of the observer. The RTE is generally written as:
\begin{equation}\label{rte}
-\cos{\theta}\frac{dI}{d\theta}=\alpha I-\Im
\end{equation}
Here we follow the notation adopted in \cite{chandra},  indicating with $I$ the light intensity (W/m$^2$) and with $\Im$ the source term, here represented by the scattering events occurring within the cloud. The absorption coefficient, $\alpha$,  is the the product of the extinction cross section, $C_{ext}$  and the number density of particles, $n$, both related to the extinction optical thickness,$\tau$, by the expression  $\tau= C_{ext} n z$ where $z$ is the geometrical thickness of the medium; finally, $\theta$ is the observation angle measured respect to the $z$ axis; if the observer is along the $z$ axis it is $\theta=0$.

To determine the radiation emerging from the cloud one has to account for both the contribution of the transmitted light along the line of sight and that of the light redirected towards the observer by scattering events. While the former contribution can be described by the LBB law, the latter is generally  more complex. In general, however, the scattered light towards the observer can be attributed 1) to single scattering events, redirecting light entering the cloud from all directions, as well as 2) to the last of two or more scattering events. Before discussing these two effects, it is useful to  make a couple of preliminary considerations as they represent limit cases of our model. 

First we notice that, thanks to the Principle of Reversibility of the optical path, for a) a perfectly isotropic diffuse source surrounding the cloud, and for b) pure dielectric particles (for which $\omega= 1$ rigorously), the probability of removing some radiation from the line of sight (due to scattering in any direction) is exactly the same as the probability that radiation is injected along the line of sight from any direction. The net result is no extinction at all. Of course, as soon as one of the two hypotheses is removed, this result is no longer valid: dropping a) we find that light scattered towards a direction from which the source is not illuminating the cloud is not compensated, while dropping b) the radiation is partially absorbed (and possibly re-emitted at other wavelengths). Our second simple consideration is that for an optically thick cloud composed by pure dielectric particles and illuminated by a source with any geometry, i.e. non isotropic, the whole radiation power entering the cloud will be isotropically scattered.

In the real world the above assumptions are hardly valid so the full RTE must be used. Our approach starts with the RTE formally expressed by Eq.~\ref{rte}, writing a convenient approximation for the source term $\Im$ that accounts for all scattering events delivering light into the line of sight. %($z$ axis in Fig. \ref{fig:figura1}).

\begin{center}
\begin{figure}[ht!]
\centering
\includegraphics[scale=0.23]{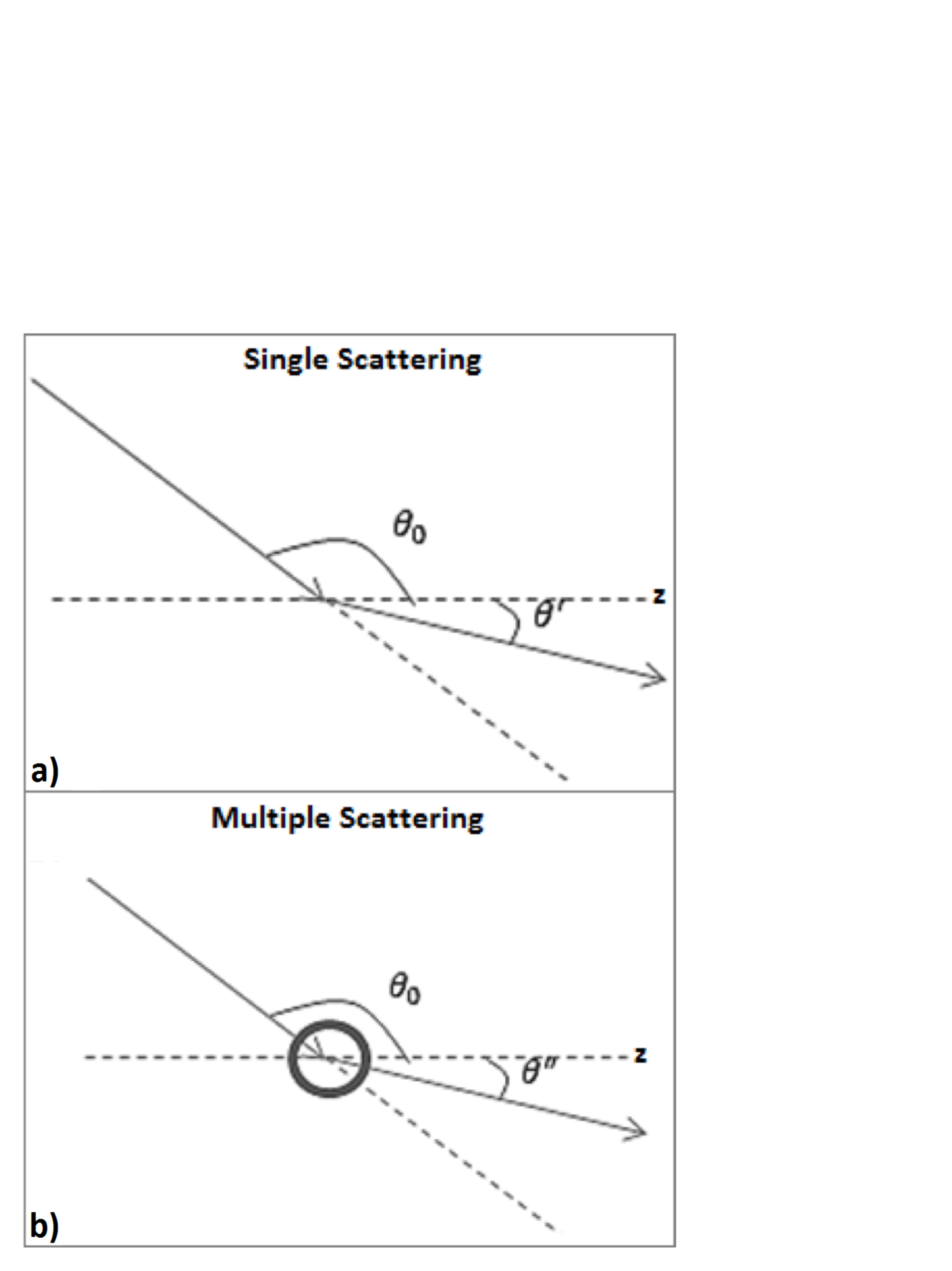}
\caption{\footnotesize Schematic representation of the single (a) and multiple (b) scattering process as considered in this work. In the case of single scattering, The radiation comes from an angular direction $\theta_0$ and it is deflected by an angle $\theta^\prime$. In the case of multiple scattering, the radiation comes from an angular direction $\theta_0$ and is deflected by the same angle as result of the last scattering event within the medium.}\label{fig:figura1}
\end{figure}
\end{center}

We assume that the cloud is a plane-parallel slab of translucent medium, with finite optical thickness $\tau$ along the $z$ direction and infinitely extended in the transversal plane. The radiation source is described by a uniform distribution of emitters, covering a given solid angle $\Omega$, and with energy spectral distribution $I(\nu)$. In Fig. \ref{fig:figura1} we put it in the negative $z$ hemisphere for simplicity, but the following arguments will not require this limitation.
 
Our expression for the source term $\Im$ in the RTE (Eq.~\ref{rte}) treats differently the single scattering and the multiple scattering events. Single scattering events can be described on the basis of the traditional scattering models for spheres \citep[see for example ][]{VdH-bohren-chandra}, or any more refined numerical and/or analytic approach, e.g. Discrete Dipole Approximation \citep{purcell} or Finite Different Time Domain \citep{FDTD}. Following \cite{chandra}, we adopt  the single scattering phase functions. Multiple scattering is described through the simplifying assumption that just two scattering events are enough to make the scattering isotropic. Multiple scattering can thus be treated as isotropic scattering minus single scattering, already accounted for in the first term. This approximation sets a limit to the reduced size of the scatterer, $\beta = k a = 2 \pi a / \lambda$,  where $a$ represents the radius and $\lambda$ the radiation wavelength, as it is well satisfied for $\beta\ll10$. A detailed discussion of this assumption, and the implications in an astrophysical context, will be presented in Section 4.2. Apart from this, we do not constrain the nature of the particles, which can be a collection of scatterers of any nature, size distribution, composition, shape, internal structure.

Given the two phase functions $p_{s}$ and $p_{m}$ for the single and multiple scattering, the RTE can be expressed as follows:
\begin{equation}\label{rte2}
\begin{split}
-\frac{dI}{d\tau}=&I_0e^{-\tau}-\\ & \: -\frac{1}{4\pi}\int_{4\pi}I_{i}\left (\tau,\mu^\prime,\varphi^\prime,\mu_0,\varphi_0\right ) p_{s}\left ( \mu^\prime,\varphi^\prime \right ) d\mu^\prime d\varphi^\prime- \\ &\:-\frac{1}{4\pi}\int_{4\pi}\tilde{I}_{i}\left ( \tau,\mu^{\prime\prime},\varphi^{\prime\prime},\mu_0,\varphi_0\right ) p_{m}\left ( \mu^{\prime\prime},\varphi^{\prime\prime}\right ) d\mu^{\prime\prime} d\varphi^{\prime\prime}
\end{split}
\end{equation}
where the polar angles $\theta$ and $\phi$ have been introduced as described in Fig. \ref{fig:figura1} and we have adopted the usual convention $\mu = \cos{\theta}$. The first term accounts for the photons injected along the line of sight from single scattering events and the second term for the photons injected through multiple scattering events.

The formal solution of Eq. \ref{rte2} takes a rather compact form:
\begin{equation}\label{sol}
I(\tau)=I_0e^{-\tau}+\frac{I_0}{2}\int_\Omega\mu_0 (1-e^{-\tau/\mu_0})\int_{-1}^{+1}P(\mu_0,\mu^\prime)d\mu_0d\mu^\prime
\end{equation}
Here $\Omega$ is the angular range subtended by the source, $I_0=\int_\Omega I_{i}\left (\tau,\mu^\prime,\varphi^\prime,\mu_0,\varphi_0\right )d\Omega=\tilde{I}_{i}\left ( \tau,\mu^{\prime\prime},\varphi^{\prime\prime},\mu_0,\varphi_0\right )$ and $P = p_{s} + p_{m}$, where $p_{s}=\frac{p(\theta)}{K}$ and $K=\int_{[0;\pi]}p(\theta)\sin{\theta}d\theta$.

To evaluate $p_{m}$, first we consider the fraction of light that is generally scattered, simply given by $1-e^{-\tau/\mu_0}$ for a medium with optical thickness $\tau/\mu_0$. Then we need to subtract the probability of having only one scattering, obtained by integrating the phase function introduced above and taking into account the extinction of the incoming radiation along the direction $\mu_0$. The phase function for the multiple scattering is then: 
\begin{equation}\label{pmult}
\begin{split}
p_{m}=1&-e^{-\tau/\mu_0}-\\ &-\frac{1}{2K}\int_\Omega\int_{-1}^{+1}\mu_0p(\mu_0,\mu^\prime)(1-e^{-\tau/\mu_0})d\mu_0d\mu^\prime
\end{split}
\end{equation}

To illustrate the results obtained from the solution \ref{sol}, we introduce the transmission as the ratio $I/I_0$.

\begin{center}
\begin{figure}[ht!]
\centering
\includegraphics[scale=0.62]{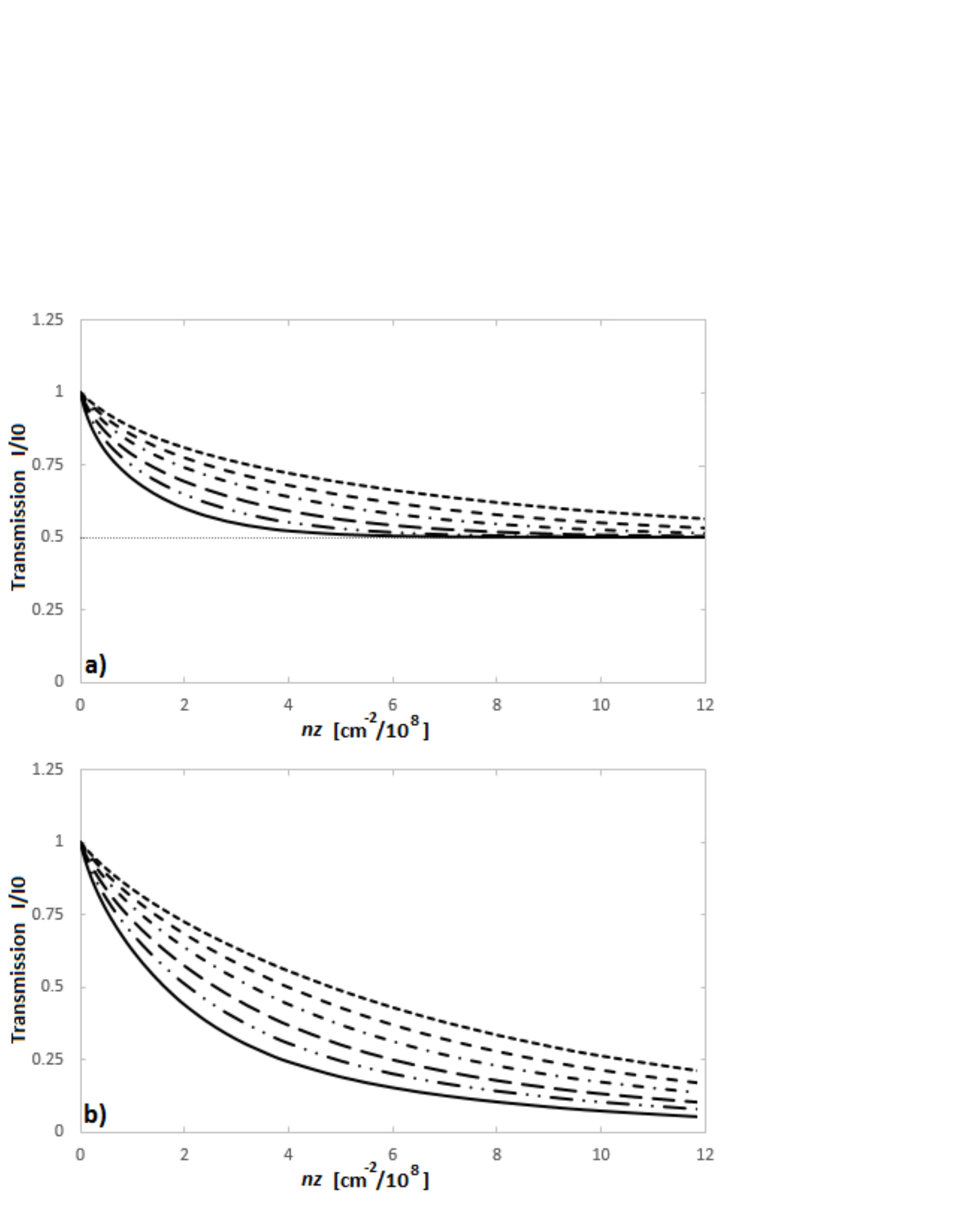}
\caption{\footnotesize RTE solutions in term of transmission ratio for radiation coming from a semi-space $(\Omega=2\pi\text{ sr})$ for different wavelengths. From top to bottom: 900, 800, 700, 600, 500 and 400 nm. (a) purely dielectric particles with diameter of 0.6 $\mu$m, refractive index $n=1.3$ in vacuum. (b) absorbing particles with the same diameter and refractive index $n=1.3+i 0.03$, giving rise to a scattering albedo $\omega_0\approx0.9$.}\label{fig:figura2}
\end{figure}
\end{center}

In Fig. \ref{fig:figura2}.a we plot an example of the transmission ratio as a function of the optical thickness (average value overall the wavelength range) of the cloud, in the ideal case of dielectric scatterers. We consider monodisperse (same size), spherical particles with refractive index $n=1.7$, diameter of 0.6 $\mu$m. The corresponding phase functions can be evaluated on the basis of the Mie expansions and the solution can be obtained by integrating over the semispace $\Omega=2\pi$ sr subtended by the background source. In this case the source has been assumed to radiate as a blackbody at 3000 K.  Different wavelengths are shown by different lines as detailed in the legend. The asymptotic behavior for large optical thickness shows that half of the light is backscattered, half is forward scattered in the cloud, consistently with what discussed above. In Fig. \ref{fig:figura2}.b we show the results obtained adding some absorption to the same particles, by adding an imaginary part \textit{i}0.03 to the refractive index. The two cases are different mostly due to the presence of a horizontal asymptote at large optical thicknesses in the case of pure dielectric particles, which disappears in the case of absorption. This is again in accordance with what we mentioned earlier invoking the Principle of reversibility of the optical path: even at the highest optical thicknesses, when the cloud completely extinguishes the incoming radiation, scattering collects radiation from other directions and delivers it along the line of sight. 

\begin{center}
\begin{figure}[ht!]
	\centering
	\includegraphics[scale=0.75]{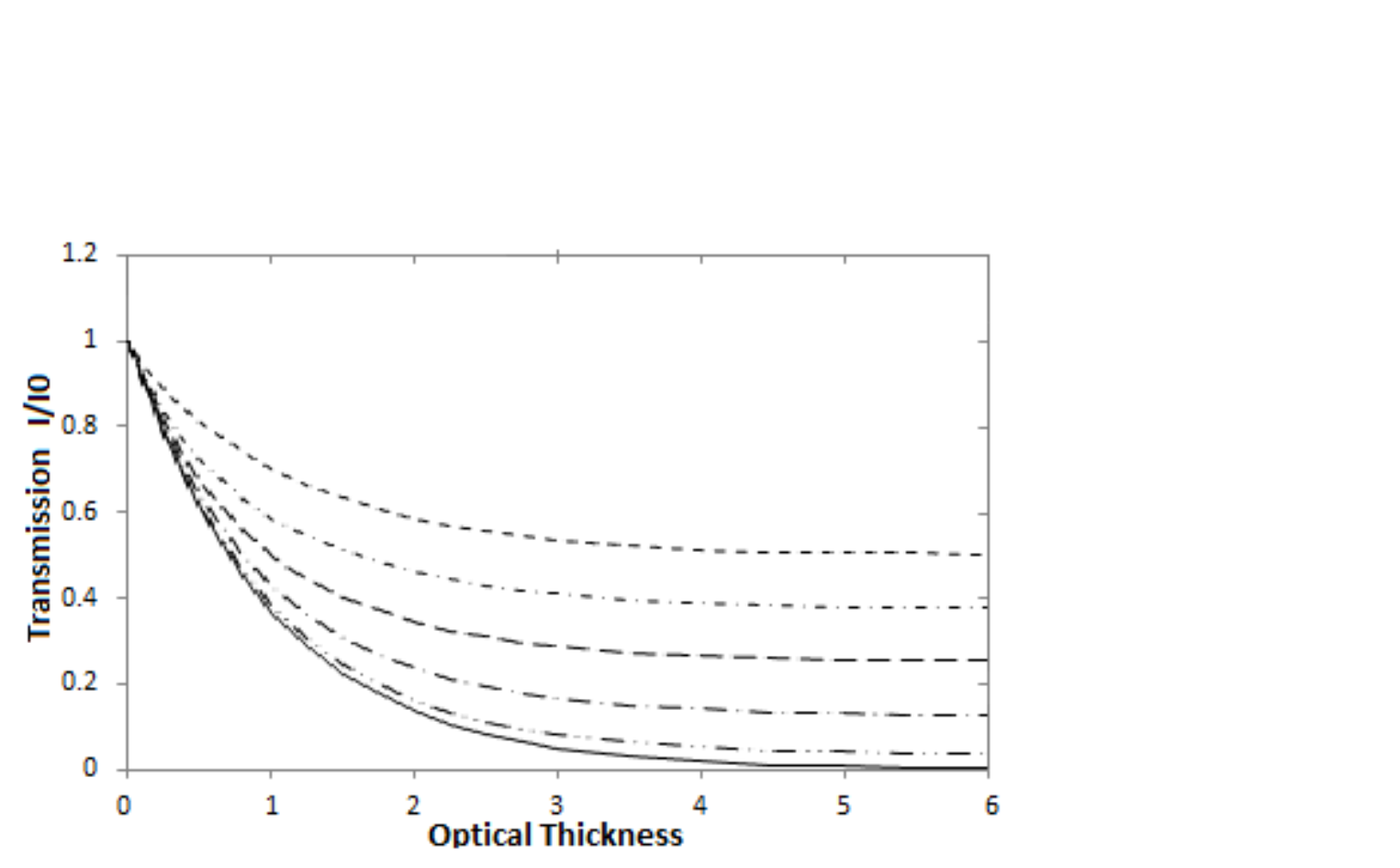}
	\caption{\footnotesize RTE solutions for different angles subtended by the source. From top to bottom the maximum illumination angles are $90^\circ$, $60^\circ$, $45^\circ$, $30^\circ$, $15^\circ$ and $0^\circ$. The LBB model (0°) gives the lowest trasmission.}\label{fig:differentiasintotiorizzontali}
\end{figure}
\end{center}

The asymptotic value of the extinction depends on the amplitude of the solid angle subtended by the source:
\begin{equation}\label{asympt}
\lim_{\tau\to +\infty}{I(\tau)}=I_0\int_D\mu_0d\mu_0
\end{equation}

where $D$ is the angular range of the azimuthal angle $\theta_0$, and $\mu_0=\cos{\theta_0}$. We have $D\subseteq [\pi/2,\pi]$.
The transmission through the cloud is systematically higher than the case of a pointlike source, where the single LBB law gives $I(\tau)/I_0=e^{-\tau}$.\\
In figure \ref{fig:differentiasintotiorizzontali} we plot the transmission ratio as a function of the optical thickness for different angular ranges $D$. We assume a source azimuthally symmetric around the direction of observation. Figure \ref{fig:differentiasintotiorizzontali} shows that there are different horizontal asymptotes for different angular ranges. Solid line corresponds to the LBB law which gives the lowest transmission.

\section{A simplified analytic solution}
In this section we derive an analytic solution of the RTE introduced above (Eq.\ref{sol}) by introducing a simplified expression for the phase functions. This solution, although approximated, can be used to evaluate the effect of the extended illumination from a few main parameters. 
\\ Let's start with the formal result for the general solutionof Eq.s \ref{rte2} and \ref{sol}. The general expression for $I(\tau)$ can be written as: 

\begin{equation}
\label{res1}
\begin{split}
I(\tau)&=I_0e^{-\tau}+\frac{I_0}{2K}\int_D\int_{[0;\pi]}\left |\cos{\theta_0}\right |\left (1-e^{-\tau/\left |\cos{\theta_0}\right |}\right )\times\\ & \qquad\qquad\qquad\qquad\qquad\times p(\theta_0;\theta^\prime)\sin{\theta_0}\sin{\theta^\prime}d\theta_0d\theta^\prime + \\ &\qquad +I_0\int_D \left |\cos{\theta_0}\right |e^{-\tau_A/\left |\cos{\theta_0}\right |}\left (1-e^{-\tau_S/\left |\cos{\theta_0}\right |}\right )\times \\ & \qquad\qquad\qquad\times \left (1-e^{-\tau/\left |\cos{\theta_0}\right |}\right )\sin{\theta_0}d\theta_0- \\ &\qquad -\frac{I_0}{2KV}\int_D\left |\cos{\theta_0}\right |\left (1-e^{-\tau/\left |\cos{\theta_0}\right |}\right )\sin{\theta_0}d\theta_0\cdot \\ & \qquad\qquad\cdot \int_D\int_{[0;\pi]}\left |\cos{\theta_0}\right |\left (1-e^{-\tau/\left |\cos{\theta_0}\right |}\right )\times \\ & \qquad\qquad\qquad\qquad\times p(\theta_0;\theta^\prime)\sin{\theta_0}\sin{\theta^\prime}d\theta_0d\theta^\prime= \\
&=I_0e^{-\tau}+\frac{I_0}{2K}\Sigma(\tau)+I_0\chi(\tau)-\frac{I_0}{2KV}\Lambda(\tau)\Sigma(\tau)
\end{split}
\end{equation}
having indicated with $\Sigma(\tau)$, $\chi(\tau)$ and $\Lambda(\tau)$ the integrals in the equation. The main obstacle to the analytic integration is represented here by the phase function $p$. We introduce therefore a simplified expression describing the diffractive behavior of the angular intensity distribution for a particle of generic size, within a given range as discussed below. The following expression for the form factors of particles with size (diameter) $d$, so that the reduced size is $b = k d / 2$, and $\beta=\frac{\pi}{b}$, exactly reproduces the zeroes of the typical diffraction patterns of objects with  diameter $d$: 

\begin{equation}
\label{res2}
p(\tilde{\theta})= \left\{
\begin{aligned} \frac{1+\cos{(b\tilde{\theta})}}{2} & \text{; }\tilde{\theta}\in [0;\beta] \\ 0 & \text{; otherwise} \end{aligned}
\right.
\end{equation}
with $p(\beta)=0$ for $\theta > \beta$. Note that this is just the first order of the typical expansion used for describing the phase functions \citep{chandra}.
\\ The normalization factor $K$ in Eq. \ref{res1} thus becomes: 

\begin{equation}
\label{res3}
\begin{split}
K&=\omega\int_{[0;\pi]}p(\tilde{\theta})\sin{\tilde{\theta}}d\tilde{\theta}= \\
&=\omega\frac{\pi^2-2\beta^2-\pi^2\cos{\beta}}{2\left (\pi^2-\beta^2 \right )}.
\end{split}
\end{equation}

This expression is valid as long as the size of the particles is small enough that $\beta \gtrsim \frac{\pi}{6}$. This is in accordance with the limited validity of the assumption based upon diffraction we make here to estimate the angular aperture of the  diffracted intensity distribution. On the other hand, also sizes much smaller than the wavelength are not well described here, since the diffraction approximation fails, and we can assume the scattered intensity to be isotropic just from the first scattering event. Nevertheless, the range of validity imposed by these arguments remains fully consistent with the aims of the model presented here, which is thought to be used for particles the size of the wavelength or slightly larger.  

By inserting the phase function \ref{res2} in Eq. \ref{res1}, we obtain the following simplified expression: 

\begin{equation}
\label{res4}
\begin{split}
\Sigma(\tau)&=\int_D\int_{[0;\pi]}\left |\cos{\theta_0}\right |\left (1-e^{-\tau/\left |\cos{\theta_0}\right |}\right )p(\theta_0;\theta^\prime)\sin{\theta_0}\times \\ &\qquad\qquad\times \sin{\theta^\prime}d\theta_0d\theta^\prime=\\&=\int_{[\pi-\delta,\pi]}\int_{[0;\pi]}\left |\cos{\theta_0}\right |\left (1-e^{-\tau/\left |\cos{\theta_0}\right |}\right )p(\theta_0;\theta^\prime)\times \\ &\qquad\qquad\times \sin{\theta_0}\sin{\theta^\prime}d\theta_0d\theta^\prime=\\&=\Bigg |\int_{[0;\delta]}d\theta_0\sin{\theta_0}\cos{\theta_0}\left (1-e^{-\tau/\cos{\theta_0}}\right )\times \\ & \qquad\qquad\times \underbrace{\int_{[\theta_0;\theta_0+\beta]}\frac{1+\cos{[b(\theta^\prime-\theta_0)]}}{2}\sin{\theta^\prime}d\theta^\prime}_{\sigma(\theta_0)} \Bigg |= \\ &= \Bigg | \int_{[0,\delta]} \sigma (\theta_0)\sin{\theta_0}\cos{\theta_0}\left (1-e^{-\tau/\cos{\theta_0}}\right ) d\theta_0 \Bigg |
\end{split}
\end{equation}
where $\pi-\delta$ is the minimum angle from which light is drawn towards the cloud, thus defining the illumination angular domain $D$. 
\\ The integral over $\theta_0$ immediately gives: 

\begin{equation}
\label{res5}
\begin{split}
\sigma(\theta_0)&=\int_{[\theta_0;\theta_0+\beta]}\frac{1+\cos{[b(\theta^\prime-\theta_0)]}}{2}\sin{\theta^\prime}d\theta^\prime= \\
&=\frac{\pi^2-2\beta^2-\pi^2\cos{\beta}}{2\left (\pi^2-\beta^2\right )}\cos{\theta_0}+\frac{\pi^2\sin{\beta}}{2\left (\pi^2-\beta^2\right )}\sin{\theta_0}
\end{split}
\end{equation}
Eq. \ref{res4} can therefore be expressed as the sum of two terms: 

\begin{equation}
\label{res6}
\Sigma_1(\tau)=\int_{[0;\delta]}\left (1-e^{-\tau/\cos{\theta_0}}\right )\sin{\theta_0}\cos^2{\theta_0}d\theta_0
\end{equation}

\begin{equation}
\label{res7}
\Sigma_2(\tau)=\int_{[0;\delta]}\left (1-e^{-\tau/\cos{\theta_0}}\right )\sin^2{\theta_0}\cos{\theta_0}d\theta_0
\end{equation}

The first of these two functions cannot be integrated analytically, while the second admits a primitive. But we notice that the two functions are almost identical, except for a scale factor which can be easily fixed by imposing the asymptotes for large $\tau$ to be equal. By evaluating the two asymptotic values analytically, we find: 
\begin{equation}
L_1=\lim_{\tau\to\infty}{\Sigma_1(\tau)}=\int_{[0;\delta]}\sin{\theta_0}\cos^2{\theta_0}d\theta_0=\frac{1-\cos^3{\delta}}{3}
\end{equation}
\begin{equation}
L_2=\lim_{\tau\to\infty}{\Sigma_2(\tau)}=\int_{[0;\delta]}\sin^2{\theta_0}\cos{\theta_0}d\theta_0=\frac{\sin^3{\delta}}{3}
\end{equation}

We end up with a simple receipt to evaluate the function $\Sigma_1(\tau)$ through the analytic expression of $\Sigma_2(\tau)$. The integration gives: 

\begin{equation}
\label{res8}
\begin{split}
\Sigma_1(\tau)&= \\
&=\frac{1}{6}\Biggl [\tau^3\Big (\text{Ei}(-\tau/\cos{\delta})-\text{Ei}(-\tau)\Big )+\\ & \:\:\:\qquad +\tau^2\left (e^{-\tau/\cos{\delta}}-e^{-\tau}\right )+2\left (1-e^{-\tau}\right )-\\ &\qquad \:\:\: -\tau\left (\cos^2{\delta}e^{-\tau/\cos{\delta}}-e^{-\tau}\right )+ \\ &\qquad \:\:\: -\cos{\delta}\left (1-e^{-\tau/\cos{\delta}}\right )\left (\cos{(2\delta)}+1\right ) \Biggr ]
\end{split}
\end{equation}
Where $\text{Ei(x)}$ is the exponential integral:
\begin{equation}
\text{Ei}(x)=\int_{-\infty}^x \frac{e^t}{t}dt
\end{equation}

\begin{equation}
\label{res9}
\begin{split}
\Sigma_2(\tau)&=\frac{L_2}{L_1}\Sigma_1(\tau)
\end{split}
\end{equation}

Eq. \ref{res4} therefore gives: 

\begin{equation}
\label{sigma}
\Sigma(\tau)=\left |\frac{\pi^2-2\beta^2-\pi^2\cos{\beta}}{2\left (\pi^2-\beta^2\right )}\Sigma_1(\tau)-\frac{\pi^2\sin{\beta}}{2\left (\pi^2-\beta^2\right )}\Sigma_2(\tau) \right |
\end{equation}

The factor $\Lambda(\tau)$ in the last term gives: 
\begin{equation}\label{laambda}
\begin{split}
\Lambda(\tau)&=\int_D\left (1-e^{-\tau/\left |\cos{\theta_0}\right |}\right )\sin{\theta_0}\left |\cos{\theta_0}\right |d\theta_0= \\
&=\frac{1}{2}\Biggl [ \tau^2\Big (\text{Ei}(-\tau)-\text{Ei}(-\tau/\cos{\delta})\Big )+\\ & \qquad\:\:\: +\tau \left (e^{-\tau}-\cos{\delta}e^{-\tau/\cos{\delta}}\right )+ (1-\cos^2{\delta})-\\ & \qquad\:\:\:-\Big (e^{-\tau}-\cos^2{\delta}e^{-\tau/\cos{\delta}}\Big ) \Biggr ]
\end{split}
\end{equation}

For dielectric particles $(\omega=0)$ the multiple scattering term in Eq. \ref{res1} brings to:
\begin{equation}\label{chhidielectric}
\begin{split}
\chi(\tau)=&\int_D\left (1-e^{-\tau/\left |\cos{\theta_0}\right |}\right )^2\sin{\theta_0}\left |\cos{\theta_0}\right |d\theta_0=\\ =&\frac{1}{2}\Biggl [ 4\tau^2\Big (\text{Ei}(-2\tau/\cos{\delta})-\text{Ei}(-2\tau)\Big )-\\ &\qquad -2\tau^2\Big (\text{Ei}(-\tau/\cos{\delta})-\text{Ei}(-\tau)\Big )+\\ &\qquad +2\tau\left (e^{-\tau}-\cos{\delta}e^{-\tau/\cos{\delta}}\right )-\\ &\qquad -2\tau \left (e^{-2\tau}-\cos{\delta}e^{-2\tau/\cos{\delta}}\right )-\\ &\qquad-2\left (e^{-\tau}-cos^2{\delta}e^{-\tau/\cos{\delta}}\right )+\\ &\qquad +\left (e^{-2\tau}-\cos^2{\delta}e^{-2\tau/\cos{\delta}}\right )+\\ &\qquad +\left (1-\cos^2{\delta}\right )\Biggl ]
\end{split}
\end{equation}

 For absorbing particles ($\omega \neq 0$) we have:
\begin{equation}\label{chhi}
\begin{split}
\chi(\tau)&=\int_D e^{-\tau(1-\omega)/\left |\cos{\theta_0}\right |}\left (1-e^{-\tau\omega/\left |\cos{\theta_0}\right |}\right )\times \\ &\qquad \left (1-e^{-\tau/\left |\cos{\theta_0}\right |} \right )\left |\cos{\theta_0}\right |\sin{\theta_0}d\theta_0=\\&=\frac{1}{4}\Biggl [-e^{-2\tau/\cos{\delta}}+e^{-\tau/\cos{\delta}}+e^{\tau(\omega-2)/\cos{\delta}}-\\ &\qquad -e^{\tau(\omega-1)/\cos{\delta}}+4\tau\cos{\delta}e^{-2\tau/\cos{\delta}}-\\ &\qquad -2\tau\cos{\delta}e^{-\tau/\cos{\delta}}-4\tau\cos{\delta}e^{\tau(\omega-2)/\cos{\delta}}+\\ &\qquad +2\tau\cos{\delta}e^{\tau(\omega-1)/\cos{\delta}}+2\tau\omega\cos{\delta}e^{\tau(\omega-2)/\cos{\delta}}-\\ & \qquad -2\tau\omega\cos{\delta}e^{\tau(\omega-1)/\cos{\delta}}-\cos{(2\delta)}e^{-2\tau/\cos{\delta}}+\\ &\qquad +\cos{(2\delta)}e^{-\tau/\cos{\delta}}+\cos{(2\delta)}e^{\tau(\omega-2)/\cos{\delta}}-\\ & \qquad -\cos{(2\delta)}e^{\tau(\omega-1)/\cos{\delta}}+8\tau^2\text{Ei}(-2\tau/\cos{\delta})-\\ & \qquad -2\tau^2\text{Ei}(-\tau/\cos{\delta})-8\tau^2\text{Ei}(\tau(\omega-2)/\cos{\delta})+\\ &\qquad +8\tau^2\omega\text{Ei}(\tau(\omega-2)/\cos{\delta})-\\ &\qquad -2\tau^2\omega^2\text{Ei}(\tau(\omega-2)/\cos{\delta})+\\ &\qquad +2\tau^2\text{Ei}(\tau(\omega-1)/\cos{\delta})-\\ & \qquad -4\tau^2\omega\text{Ei}(\tau(\omega-1)/\cos{\delta})+\\ &\qquad +2\tau^2\omega^2\text{Ei}(\tau(\omega-1)/\cos{\delta})\Biggr ]-\\ &\:\:\:-\frac{1}{2}\Biggl [-e^{-2\tau}+e^{-\tau}+e^{\tau(\omega-2)}-e^{\tau(\omega-1)}+2\tau e^{-2\tau}-\\ & \qquad -\tau e^{-\tau}-2\tau e^{\tau(\omega-2)}+\tau e^{\tau(\omega-1)}+\tau\omega e^{\tau(\omega-2)}-\\ & \qquad-\tau\omega e^{\tau(\omega-1)}+4\tau^2\text{Ei}(-2\tau)-\tau^2\text{Ei}(-\tau)-\\ & \qquad -4\tau^2\text{Ei}(\tau(\omega-2))+4\tau^2\omega\text{Ei}(\tau(\omega-2))-\\ &\qquad -\tau^2\omega^2\text{Ei}(\tau(\omega-2))+\tau^2\text{Ei}(\tau(\omega-1))-\\ & \qquad -2\tau^2\omega\text{Ei}(\tau(\omega-1))+\tau^2\omega^2\text{Ei}(\tau(\omega-1))\Biggr ]
\end{split}
\end{equation}

Finally, the term $V$ is simply obtained from the angular domain of illumination to be $1/2 \sin^2\delta$. This completes the analytic integration of the general solution of Eq.~\ref{res1} for $I(\tau)$ in terms of the particle diameter, $d$, and the angular extent of the source, $\delta$. By inserting $\beta=\frac{\pi}{b}$ and $\delta$ in $\Sigma(\tau)$, $\chi(\tau)$, $\Lambda(\tau)$ (Eq.s \ref{sigma}, \ref{chhi}, \ref{laambda}) and $K$ (Eq.~\ref{res3}) one obtains the analytic expression for $I(\tau)$ from Eq.~\ref{sol}.

\begin{center}
\begin{figure}[ht!]
\centering
\includegraphics[scale=0.7]{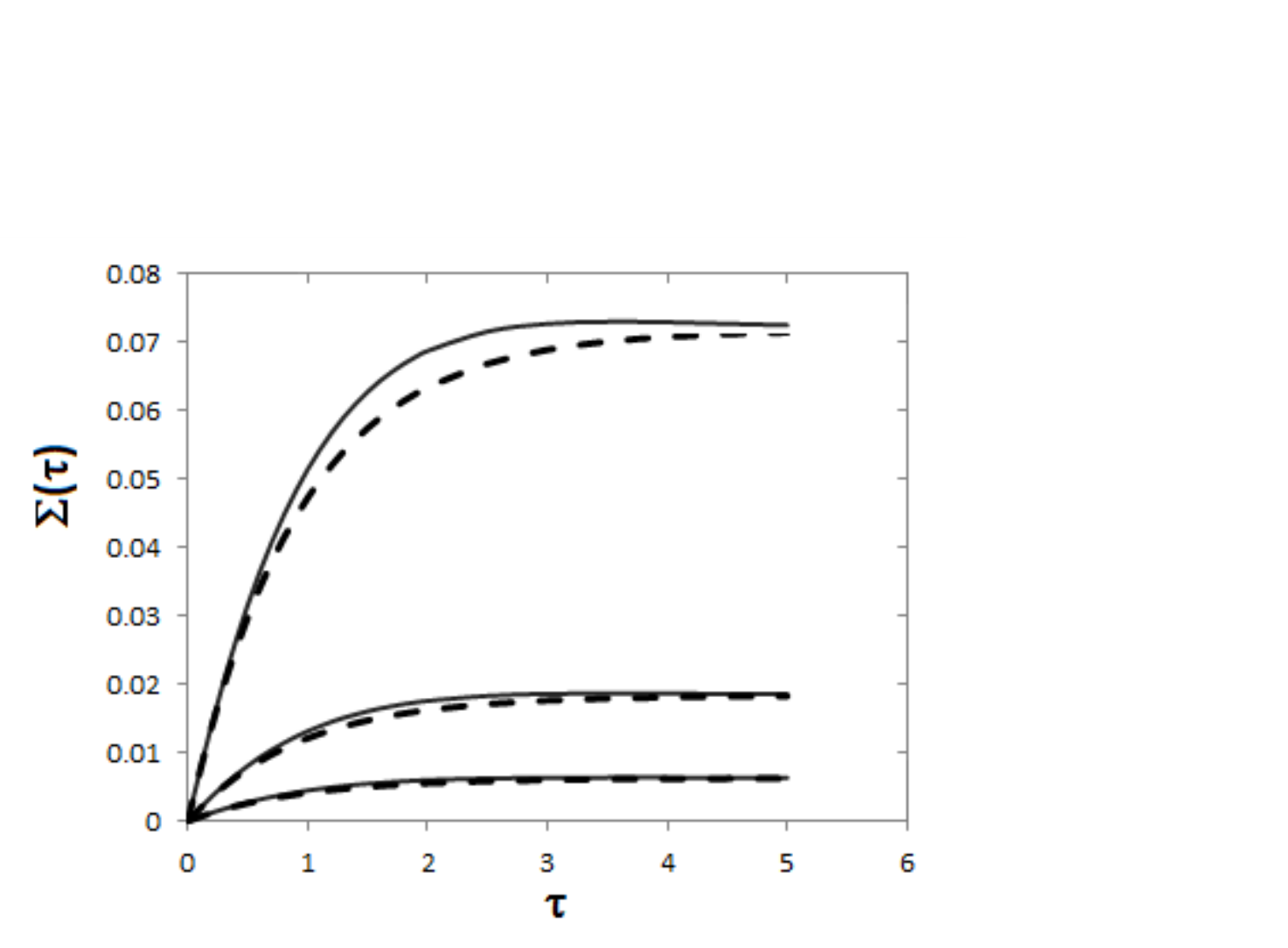}
\caption{\footnotesize Comparison of the functions $\Sigma(\tau)$ in Eq. \ref{sigma} to the functions obtained through  numerical integrations. Three $\beta$ values are considered here, 0.25, 0.60, 1.66 rad from top to bottom. Solid lines represent the analytic solutions, dashed lines the numerical approximations.}
\label{fig:SigmafunzTauvarieBeta}
\end{figure}
\end{center}

Here below we first compare the results of the analytic integrations leading to the expression for $\Sigma(\tau)$, compared to the results obtained through numerical integration of the same functions. This allows us to illustrate the range of validity of the assumptions made to overcome the non--integrability in the expression bringing to $\Sigma_1$. In Fig. \ref{fig:SigmafunzTauvarieBeta} the results are compared as a function of $\tau$ for various $\beta$. The accordance is good for $\beta > \frac{\pi}{6}$ approximately. The highest discrepancy, about 10\%, is at $\tau\approx 1$.

We also compared two transmission curves derived with the analytic expression obtained here using the simplified phase function and the numerical integration of Eq. \ref{sol} with the correct phase function obtained from the Mie theory. The results show that the analytic solution provides an excellent approximation to the extinction curve with a maximum discrepancy of 2\%.

The condition $\beta>\frac{\pi}{6}$ ultimately corresponds to $b<\frac{6\lambda}{\pi}\approx 2\lambda$. This assumption is fully consistent with the basic assumption of our model, which requires the particle to be small enough to make the light scattered twice (or more) isotropic.

%\begin{center}
%\begin{figure}[!h]
%\centering
%\includegraphics[scale=0.7]{Luce_trasmessa_analitica_e_doppia3.pdf}
%\caption{\footnotesize Comparison between the analitical solution (solid lines) and the numerical integration (dashed lines) evaluated for three $\beta$ values: 0.25, 0.60, 1.66 rad from top to bottom. Solid lines represent the analytic solutions, dashed lines the numerical approximations.}
%\label{confronto}
%\end{figure}
%\end{center}

\section{Experimental validation}
\begin{center}
\begin{figure}[ht!]
\centering
\includegraphics[scale=0.3,angle=-90]{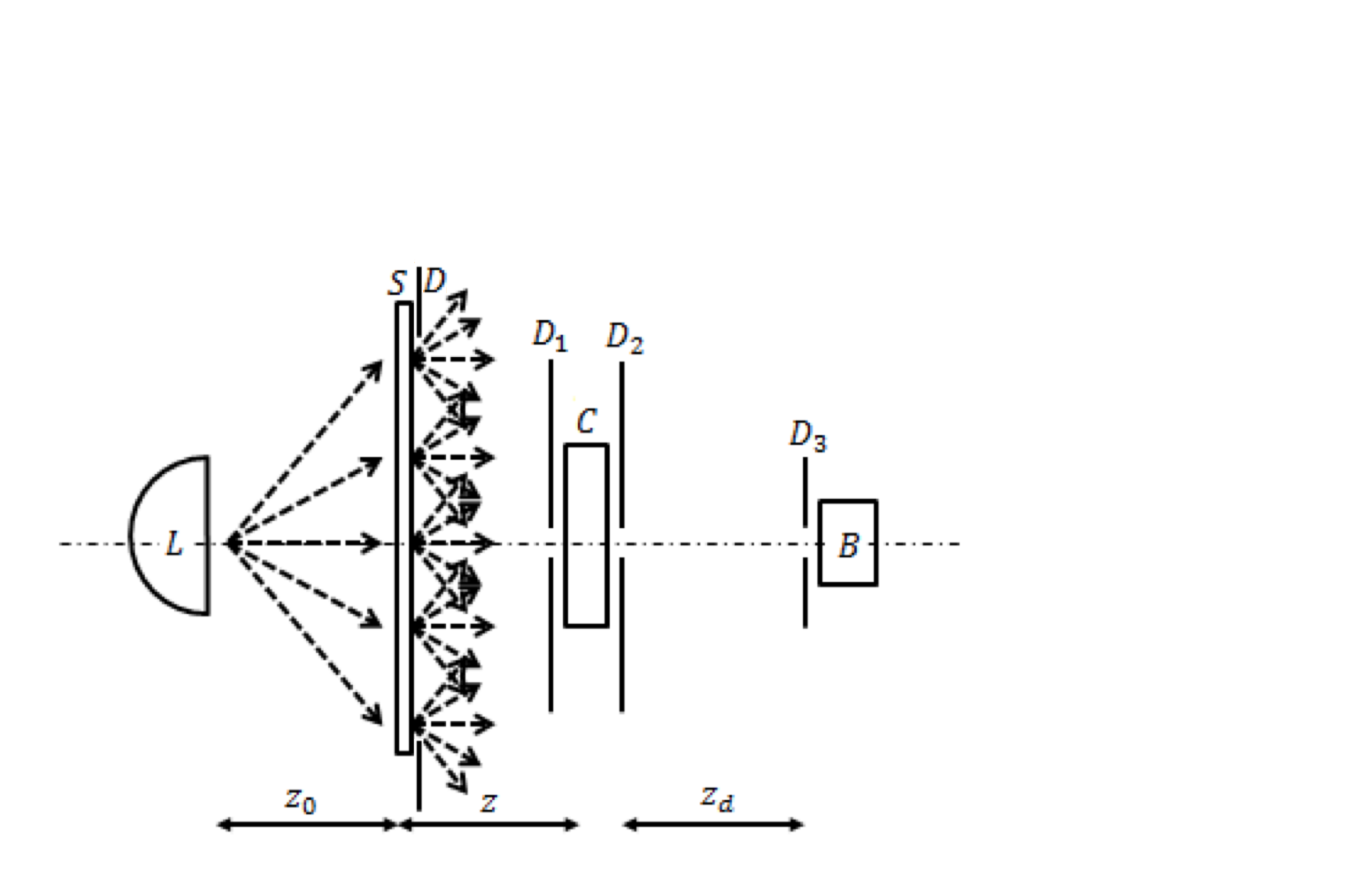}
\caption{\footnotesize Schematic of the experimental apparatus. $L$: halogen lamp; $D$: large diaphragm; $S$: diffuser acting as the diffuse extended source; $C$ sample cell; $D_1$, $D_2$, $D_3$: diaphragms; $B$: bolometer. Dashed arrows schematically represent the directions of the light rays emerging from the lamp and the diffuse source.}\label{fig:figura3}
\end{figure}
\end{center}

Even if our approximate solutions closely match the results of a rigorous treatment, given the number of assumptions we decided to compare our findings against real data. In this section we present the experimental apparatus we used to validate our results. 
\\The set-up is schematically sketched in Fig. \ref{fig:figura3}. A halogen lamp L (USHIO mod. EKE; power 150 W) illuminates a white diffuser acting as the source, $S$, composed by a 1 mm thick layer of titanium oxide deposited onto a glass surface, delimited by a circular diaphragm $D$ (50 mm in diameter, or smaller). The optical axis is defined by the position of the source and the center of the diffuser; the diffuser is aligned perpendicular to the optical axis at a distance $z_0 = 56.5$ mm from the light source. A fraction of the diffused light impinges onto a cell, $C$, uniformly filled with a water suspension of particles, which represents the scatterer system undergoing the radiative transfer phenomena to be characterized. The cell is placed on a distance $z$ from the diffuser, setting the solid angle $\Omega$ from which light illuminates the sample; by varying $z$ we can obtain different illumination conditions. A  second diaphragm, $D_1$ (2 mm in diameter), is placed just before the cell to prevent backscattered light to impinge onto the diffuser and then onto the cell again. Downstream the cell, a couple of small diaphragms, $D_2$ and $D_3$ (2 mm in diameter), separated by a distance $z_d$ selects a very narrow set of directions just around the optical axis. A bolometer (mod. IF PM from \textit{Industrial Fiber Optics})%\citep{bolometro}
measures the integrated power %of the selected light 
across the entire wavelength range.

\begin{figure}[ht!]
\centering
\includegraphics[scale=0.65]{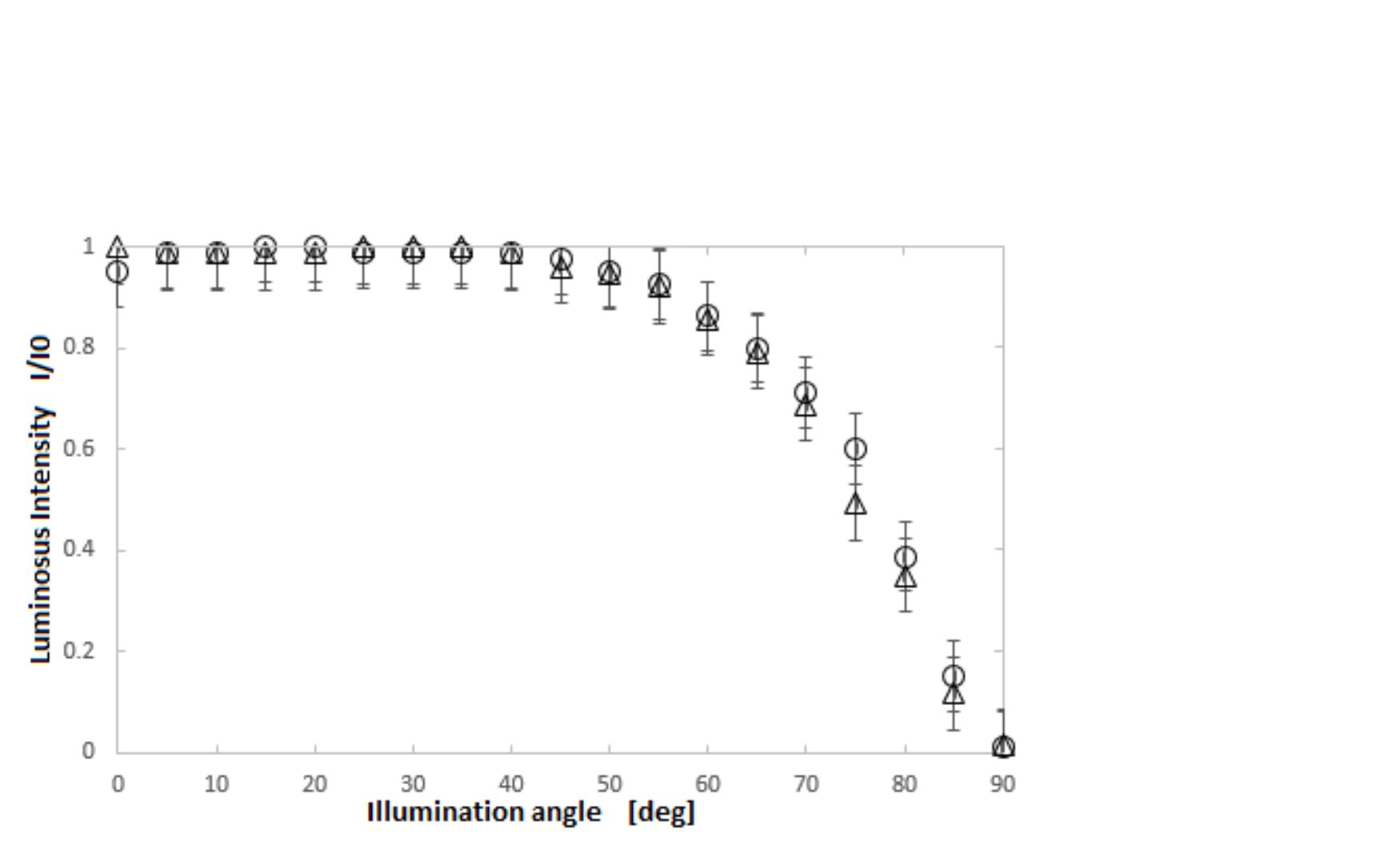}
\caption{\footnotesize Angular dependence of the intensity emitted by the extended source and impinging onto the cell, measured at distances $z=440$ mm (circles) and $z=340$ mm (triangles).}\label{fig:figura4-6}
\end{figure}

Out methodology is as follows. First, the angular profile of the light directed towards the cell is  measured to characterize the source. Then, by removing the diaphragm $D$ and with the cell filled of pure water, the bolometer is placed onto a goniometric rotation stage and an accurate characterization of the intensity passing through the cell is done. The results are represented in Fig.\ref{fig:figura4-6}. %https://preview.overleaf.com/public/cqvqfjfggzsk/logs/785b97483a4629c0039081544f995002.log

%\begin{center}
In figure \ref{fig:figura4-6} we present the intensity of the light emitted by the diffuser. By varying the distance between diffuser and bolometer, we obtain that the light intensity is uniform within 5\% up to an angle of $45^\circ$. Notice the excellent reproducibility of the results for different distances (circles and triangles, see caption). \\ 
%\end{center}
%\begin{center}
\begin{figure}[ht!]
\begin{center}
\centering
\includegraphics[scale=0.66]{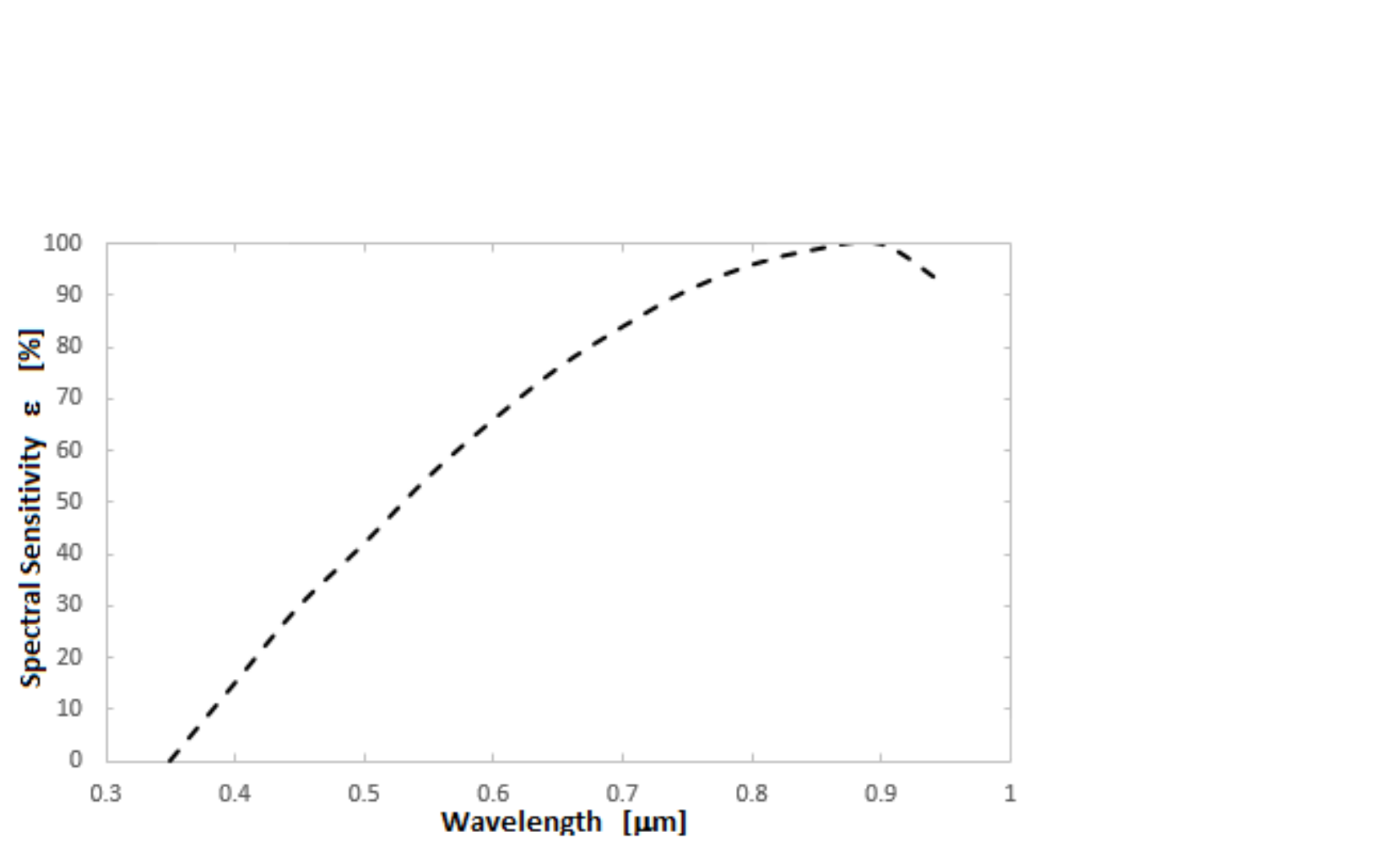}
\caption{\footnotesize Spectral sensitivity $\varepsilon$ of the bolometer.}\label{fig:figura5}
\end{center}
\end{figure}
%\end{center} 

We must also take into account the spectral sensitivity of the sensor used in our measurements, reported in Figure \ref{fig:figura5}\footnote{datasheet available at \textit{http://i-fiberoptics.com/laser-accessories-detail.php?id=1018})}, and the spectral intensity of the illuminating lamp, a black body at temperature $T=3000$ K (see above). Hereafter we will refer to data corrected for both effects.

In the following sections we report the results obtained with $z=44$ mm, corresponding to an angular aperture of the diffuse source of $21\degr$, in two cases: a suspension of calibrated particles and a set of non-homogeneous samples.\\ \\

\subsection{Transmission through collections of monodisperse, calibrated samples}
Here we present the results obtained using water suspensions of calibrated, monodisperse spherical latex particles. For these particles the whole set of parameters is well known in terms of particle size, shape, composition and concentration. Therefore the radiative transfer model can be solved without any free parameter. Scattering properties have been evaluated with a consolidated code \citep{Lompado2002}. We used the approximated value for the refractive index $n=1.59$ overall the range of wavelengths. This is the average value of the refractive indexes within the spectral range we consider here \citep{Ma2003}. Notice that the change in refractive index is small, ranging from 1.61 (400 nm) to 1.575 (900 nm), while the imaginary part is negligible.

\begin{center}
\begin{figure}[ht!]
\centering
\includegraphics[scale=0.65]{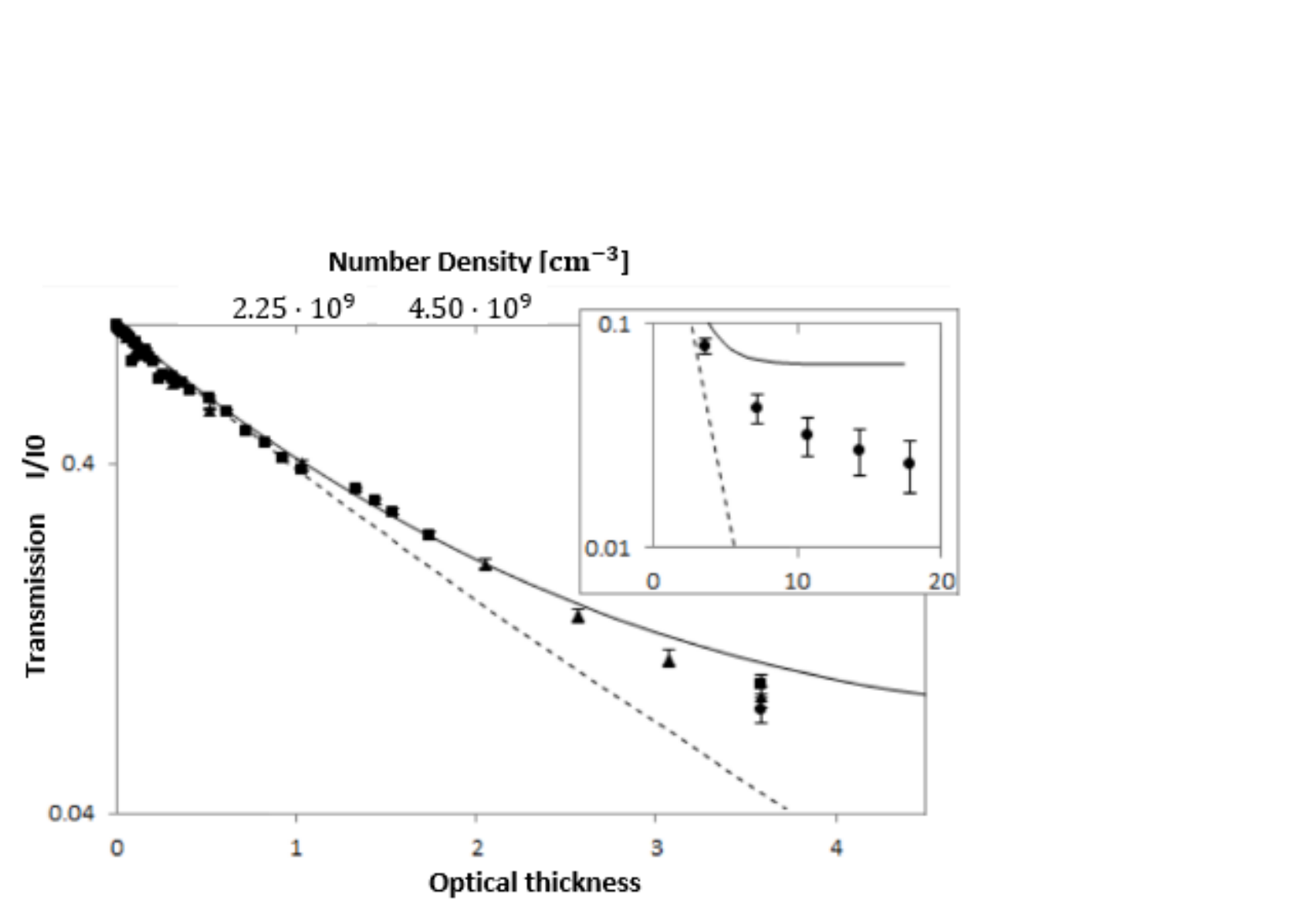}
\caption{\footnotesize Experimental results obtained with calibrated, monodisperse polystyrene spheres. Dashed line represents the normalized transmission for collimated light without any multiply scattered component. Continuous line represents the result obtained with our model (without free parameters). In the inset the data obtained at the highest densities show a deviation from the results of the model.}\label{fig:figura6}
\end{figure}
\end{center}

The radiative transfer in conditions of high optical thickness depends strongly on the  angular source extension. Since in this case the light propagates in completely random directions, the solid angle subtended by the source determines the amount of light emerging in any given direction.
Measurements performed by changing the diameter of the diaphragm limiting the size of the source $S$, thus changing the angular aperture of the illuminating source, provide results in agreement with Eq.~\ref{asympt}.

\subsection{Polydisperse, nonspherical particles: effects of absorption}
Here we drop the assumptions of well characterized monodisperse and spherical particles, moving to the study of fine powders with and without absorption. This covers all the cases of interest for interstellar grains.% \citep{Ricci}.\\
If the real part of \textit{n} is large, the grain is an effective scatterer, which is the case for dielectric grains or icy grains.\\
If the imaginary part is large, the grain is an effective absorber, e.g., as is the case for metallic grains.
 %In order to solve the model without any free parameter.
\\ We have characterized each sample by performing independent ancillary measurements. As detailed in the Appendix, we used the following tools: 1) an optical turbidimeter, to measure the optical thickness; 2) absolute measurements of small angle light scattering, to get the angular phase function and the total scattering cross section, 3) spectrophotometry, to estimate the dependence of the parameters on the wavelength.

We performed a variety of measurements on two types of samples: 1) water suspensions of Cerium oxide, composed by compact grains with negligible absorption, and 2) black carbon, which is endowed with high absorption at all wavelength and a very fluffy grain structure. Care has been taken to properly handle the samples to avoid sedimentation, aggregation, etc. Also, typical aggregation time constants have been evaluated and the measurements have been performed on the shortest possible time scale to reasonably exclude the formation of clusters. Independent checks have been done to verify the stability of the optical properties during the measurements.

\begin{center}
\begin{figure}[ht!]
\centering
\includegraphics[scale=0.5]{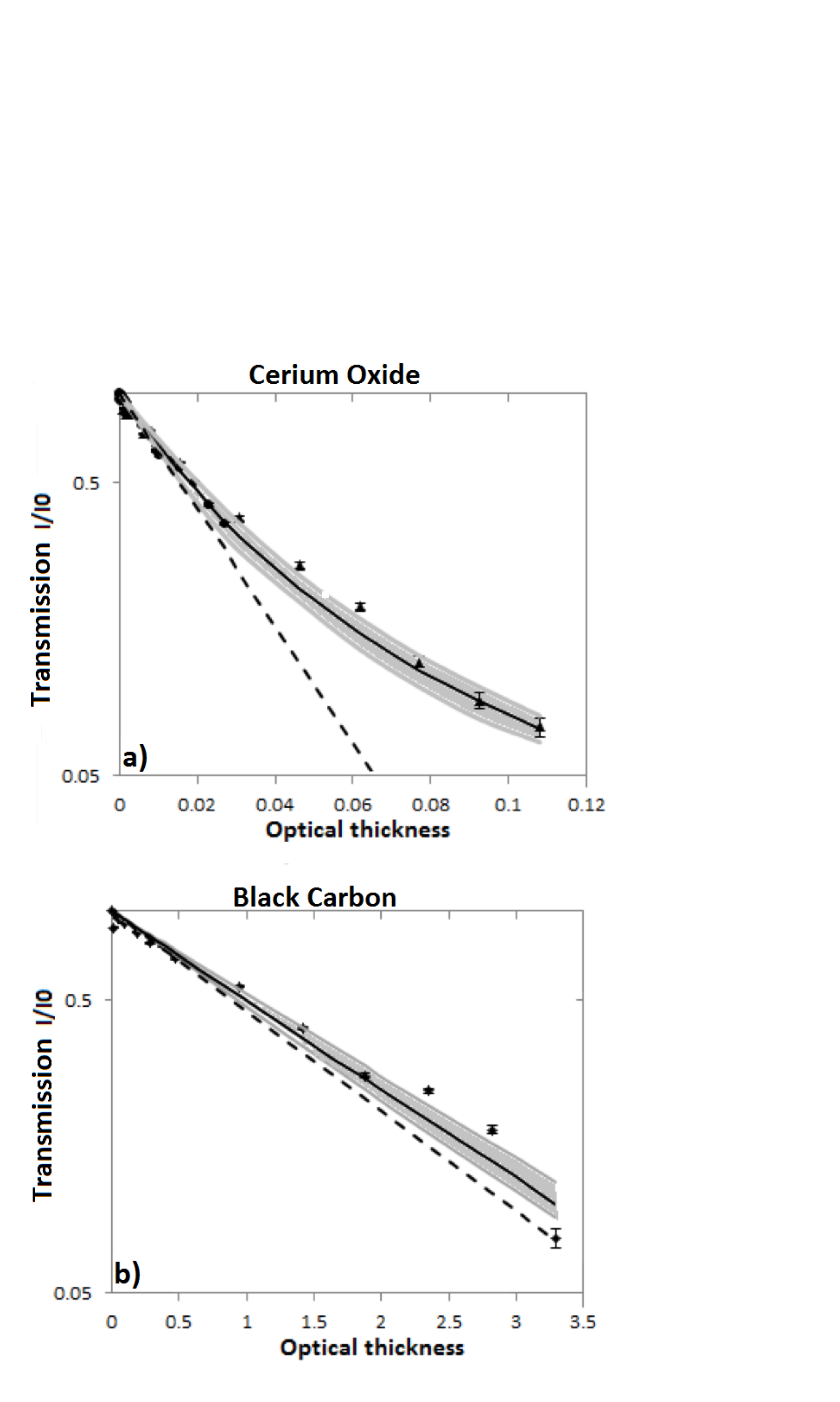}
\caption{\footnotesize Experimental results obtained with suspensions of ceria oxide (a) and black carbon (b). Dashed line represents the normalized transmission for collimated light. Continuous line represents the result obtained with our model (without any free parameters). The region evidenced in grey indicates the uncertainty introduced in the model by exaggerating the uncertainties in the phase function.}\label{fig:figura8}
\end{figure}
\end{center}

In Fig. \ref{fig:figura8}a) we plot the experimental results obtained for the normalized transmission through a suspension of Cerium oxide at different concentrations. Samples have been prepared by diluting a suspension 5\% concentrated by mass. The same sample has been accurately characterized as described in \cite{Potenza2015}. Results are plotted versus the optical thickness obtained from the turbidimeter. The deviation from the collimated beam (dashed line) is similar to the previous case. Having measured all the optical properties necessary for modeling the radiative transfer, we get nice agreement between our model and the data without recurring to any free parameter.

In Fig. \ref{fig:figura8}b) we plot the normalized transmission through a suspension of black carbon as a function of the optical thickness, evaluated again as described above. Suspensions of dry powder produced by Sennelier, one of the most famous dry pigment, have been used. They have been prepared by suspending dry powder in water, passing them in ultrasound bath until the properties of the suspension stabilized \citep{Sanvito}, and finally diluted for the measurements.
For the optical properties of this material we refer to \cite{campbell1999}, \cite{Bergstrom2006} and to direct measurements performed on the sample considered here \cite[see the Appendix and][]{Sanvito}.  The deviation with respect to the collimated beam is smaller than for dielectric particles, albeit non negligible. Moreover, unlike the results plotted in Fig. \ref{fig:figura8}, the asymptotic behavior at high optical thickness makes the transmission vanishingly small, as expected.

\section{Application to astronomical grains}
In this last section we discuss the effect of diffuse illumination on the spectral extinction when the cloud of particles has composition and size distribution similar to those found in astrophysical cases;  first we consider the classic grains of \cite{Draine}, then a mixture of grains representative of protoplanetary disks, with a special application to the case of the giant dark silhouette disk 114-426 in Orion.

\begin{center}
\begin{figure}[ht!]
\centering
\includegraphics[scale=0.5]{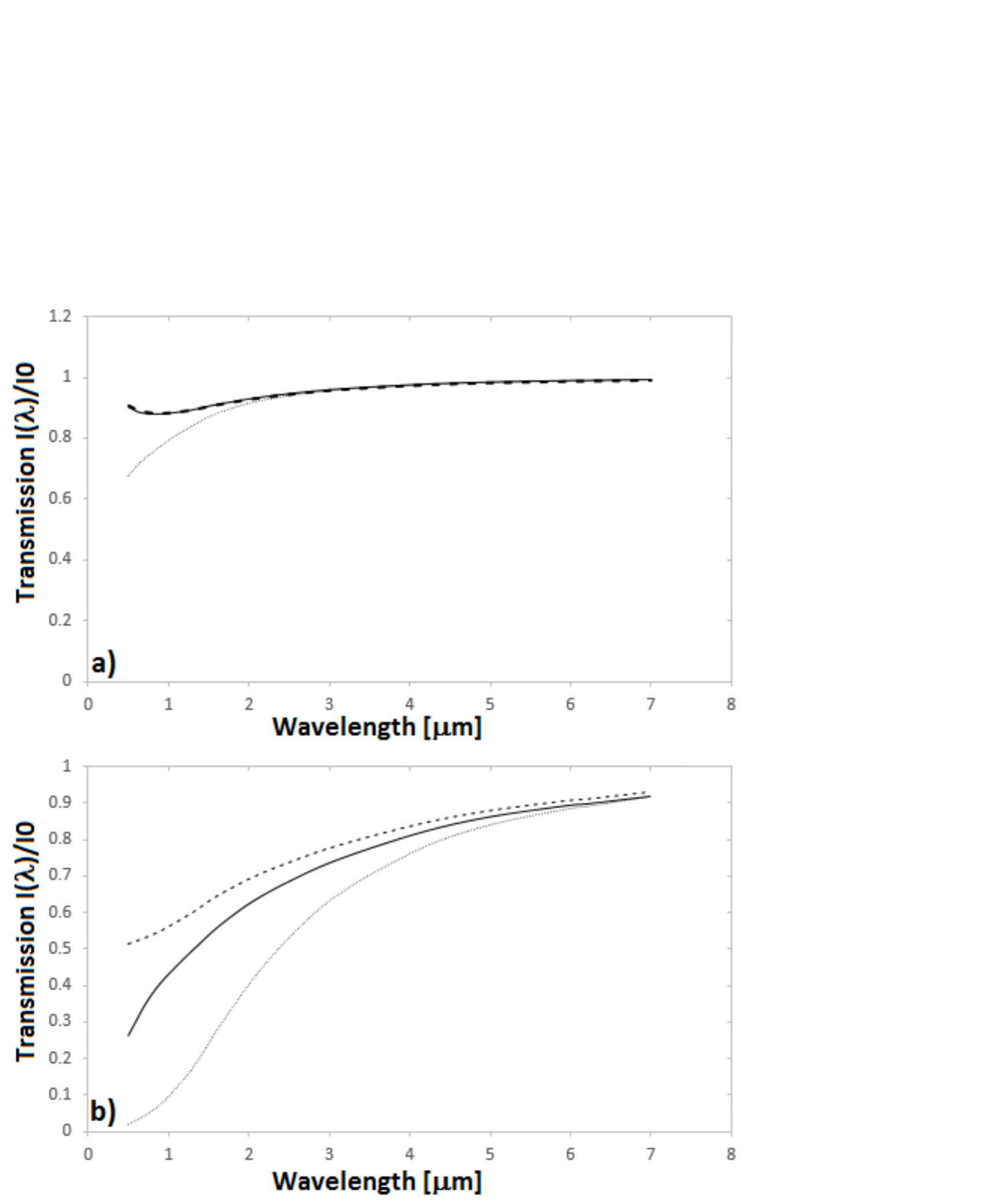}
\caption{\footnotesize Diffuse transmission in the range 400 nm - 8 $\mu$m through a cloud of interstellar dust as in \cite{Draine}, for an optically thin ($\tau=0.5$ top panel) and optically thick ($\tau=5$, bottom panel) case. The curves refer to the present method (solid lines), compared to the usual LBB model (dotted lines) and zero absorption (dashed lines) grains, treated with our method.}\label{fig:figuraDrainearticolo}
\end{figure}
\end{center}

\cite{Draine} provide a classic recipe for interstellar dust based on a combination of silicates and graphite. For our calculations we adopt the real and imaginary parts of the refractive index of their mixture. We assume a grain 
%accordingly to \citep{Ricci}, i.e. a 
size distribution $n(a) \propto a^{-q}$, with $a$ the particle radius and $q$ the spectral index. We set $q = 3.5$ and a size range from $a_{min} = 0.1$ $\mu$m to $a_{min} = 0.5$ $\mu$m
%The same size distribution of above and the approximated model for the scattering and absorption have been used. 
\\ The top panel of Fig. \ref{fig:figuraDrainearticolo} shows the $\tau$ = 0.5 case over the visible and IR wavelength range up to 8 $\mu$m. Large differences arise in the visible range, as the scattering efficiency, decreasing with wavelength, has negligible effect in the IR. By contrast, in the visible, where the reduced size of the particle, $\beta =  ka$, is close to unit, diffuse transmission is heavily affected by the presence of an extended background source, changing the reddening effect expected on the basis of the LBB law into an almost uniform transmission. For large optical thickness, $\tau = 5$ (bottom panel),   the visible range is completely dominated by scattered light, and the wavelength dependence of the transmitted light (appearing as reddening) is appreciably shifted to shorter wavelengths with respect to the LBB model. In this case the presence of an extended source affects the reddening curve even in the IR region, and  the transmission remains generally higher with respect to the LBB law.

%Two examples of the results of the present method (solid lines) compared to the usual LBB model (dotted lines) for ? = 0.5 (a) and 5 (b) for wavelengths in the range 400 ? 1000 nm. Dashed lines represent the results our model for zero absorption. Transmission through a cloud of particles composed by silicates, carbonaceous, ice and vacuum has been assumed accordingly to Ricci et al. (2010). The effect of the diffuse illumination of the cloud is evident and gives rise to appreciable changes in the shape of the transmission spectrum. In (a) no reddening is expected, at variance with the LBB law. In (b) the wavelength dependence giving rise to reddening is appreciably different from the LBB expectations. The influence of absorption is also evident. Notice that for ? = 5 the light intensity is mainly due to light from the diffuse source which is scattered towards the observation direction.

For protoplanetary disk grains, we use the set of refraction indexes provided by \cite{Ricci} and the grain composition adopted in \cite{Pollack} (30$\%$ vacuum, 21$\%$ carbonaceous, 7$\%$ astronomical silicates, 42$\%$ water ice). We maintain the size distribution $n(a) \propto a^{-3.5}$ adopted for the \cite{Draine} grains.
By evaluating the complex dielectric functions for each material from the tabulated refractive indexes, and by using the mean field approximation for the composite grains \citep{BohrenHuffmann} we obtain the grain refractive index as a function of wavelength $\lambda$. \\
% \citep[as adopted e.g. by ][]{Miotello et al}. 
If we assume the particles to be spherical, a reasonable approximation in our case as discussed by \cite{Pollack}, we can use the analytic approach described earlier by
adopting the reduced size of the scatterers (scattering form factor) 
 %inserting the scattering form factor defined as in Eq. \ref{res2}, 
 %
$\beta (\lambda) = 2\pi a / \lambda$. For the sake of simplicity, the size range is divided into 5 bins 0.1 $\mu$m wide each, and the optical properties calculated for the corresponding size values $a_i$, $i=1,...,5$. 
In order to maintain the analytic approach presented so far, we adopt an approximate expression for evaluating the cross section and the single scattering albedo following \cite{VdH-bohren-chandra} (chapter 2); analytic expressions for the extinction and absorption cross sections allow us to derive the scattering term.\\
According to the notation used in \cite{VdH-bohren-chandra} (the refractive index is described as $m=n-in^\prime$), we obtain the scattering albedo and optical thickness in terms of the absorption and extinction \textit{efficiency factors}: \begin{equation}\label{eq:Qvaluesextandabs}
\begin{aligned}
& Q_{ext}(a,\lambda)=\frac{C_{ext}}{\pi a^2}=4\text{Re}[K(i\rho+\rho\tan{\gamma})] \\
& Q_{abs}(a,\lambda)=\frac{C_{abs}}{\pi a^2}=2K(4\beta n^\prime)
\end{aligned}
\end{equation}
where $\rho=2\beta (n-1)$ and $\tan{\gamma}=\frac{n^\prime}{n-1}$ and:
\begin{equation}\label{eq:funzioneKperQvalori}
K(z)=\frac{1}{2}+\frac{e^{-z}}{z}+\frac{e^{-z}-1}{z^2}
\end{equation}
We obtain the scattering albedo as:
\begin{equation}\label{eq:scatteringalbedo}
\omega(a,\lambda)=1-\frac{Q_{abs}(a,\lambda)}{Q_{ext}(a,\lambda)}
\end{equation}
The approximations leading to Eq.\ref{eq:Qvaluesextandabs} are valid for the goal of this case study, as we are aiming at a quantitative comparison of the diffuse transmission calculated with our model against the basic LBB expectations. Higher accuracy can be achieved by introducing the full Mie expansions.
With these elements, the extinction ratio of the cloud $r = I / I_0$ is obtained as a function of wavelength $\lambda$. Here we assumed the cloud to be illuminated from the full hemisphere opposite to the observer $(\delta = \pi/2)$.

\begin{center}
\begin{figure}[ht!]
\centering
\includegraphics[scale=0.5]{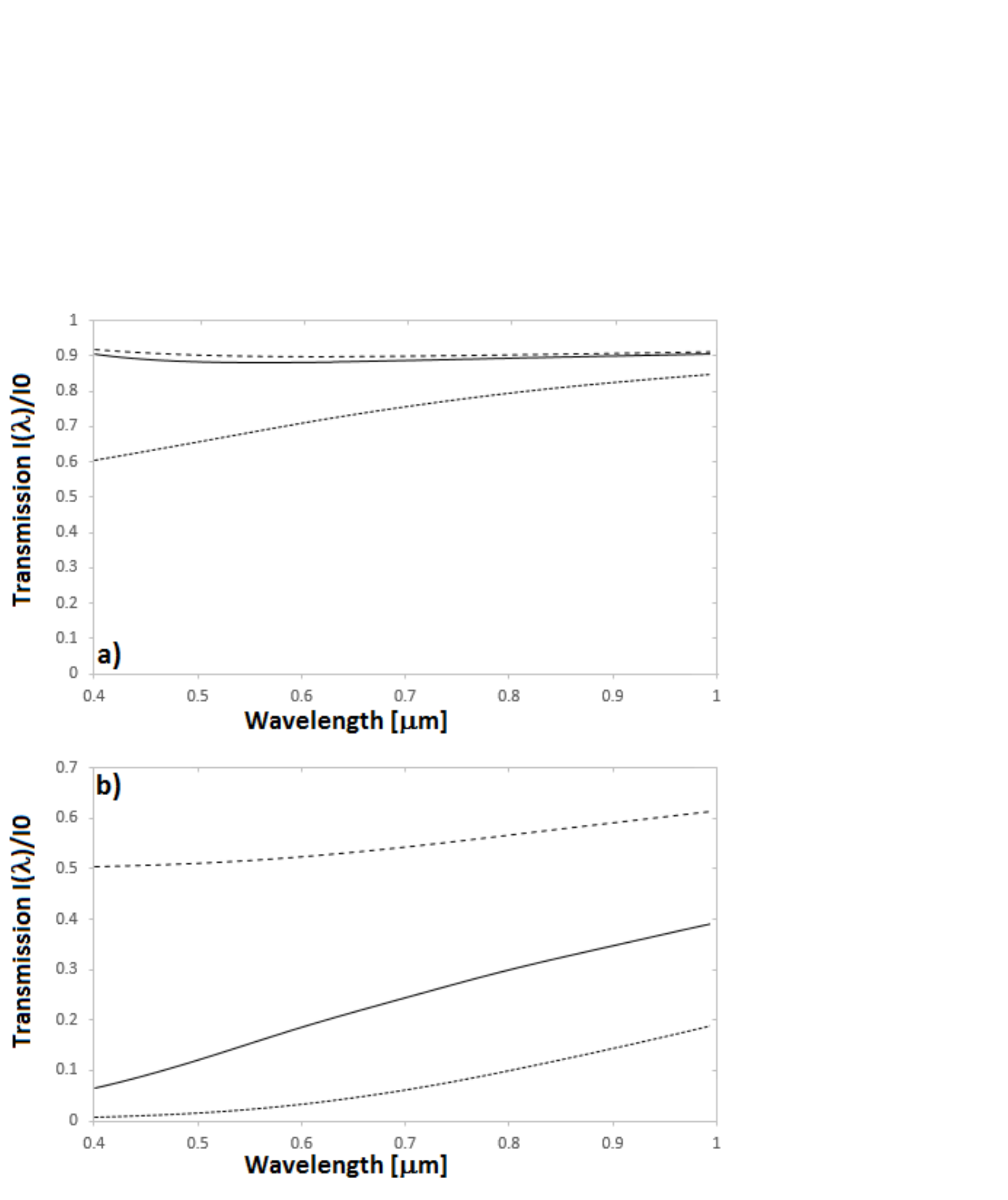}

\caption{\footnotesize Similar to Figure~\ref{fig:figuraDrainearticolo} for grains typical of protoplanetary disks, according to \cite{Ricci}, in the wavelength range 400 $-$ 1000 nm. Solid lines: our model with uniform background illumination; dotted lines: LBB model; dashed lines: dielectric (zero absorption) grains treated with our model.}
\label{fig:NewFigura10articolo}
\end{figure}
\end{center}

Fig. \ref{fig:NewFigura10articolo} shows the diffuse transmission estimated using our model for a cloud of dust composed by a mixture of silicates, carbonaceous, ice and vacuum as described in \cite{Ricci} for two values of the cloud optical thickness, $\tau$ = 0.5 (top panel) and $\tau$ = 5 (bottom panel). For clarity, we are referring here to the maximum value of $\tau$ over the considered wavelength range. The effect of the diffuse illumination of the cloud is evident, giving rise to appreciable changes in the shape of the transmission spectrum. When $\tau=0.5$, the departure from the LBB law makes  the extinction nearly wavelength. When $\tau=5$, the wavelength dependence originates a reddening law appreciably different from the one resulting from the LBB treatment. The influence of absorption is also evident. Notice that for $\tau$ = 5 the light intensity is mainly due to light from the diffuse source which is scattered towards the observer.

%In Fig. \ref{fig:newfiguare10araticolo}.a), obtained for $\tau$ = 0.5, 
The top panel shows
that in the optically thin case there are clear differences between the reddening effect expected on the basis of the LBB law (dotted line) and the nearly uniform extinction predicted by our model (solid line). The dashed line represents the result obtained by artificially forcing the absorption to zero: the effect of absorption is negligible.
%\\ In (b), $\tau$ = 5, things are completely different.  %, although appreciably shifted at lower wavelengths in the present case with respect to LBB law. 
The bottom panel shows how things change in the optically thick case.
Over the visible spectrum most of the light comes from scattering instead of being passed undisturbed through the cloud, as evidenced by the negligible values obtained for the LBB model at the shortest wavelengths compared to the finite intensity obtained with our model. The presence of scattering increases the measured transmission with respect to the LBB law, but the reddening effect is present in both cases.
If we artificially set the absorption to zero, reddening becomes negligible as the light coming from the diffuse source is dominated by scattering. 
This  case of non-absorbing material, although slightly unphysical here, warrants a final remark. For a cloud with large optical thickness with no absorption, isotropic scattering guarantees that half of the total amount of light emerges towards the observer. This is in accordance with the limit case we discussed in Section 2.

Finally, to assess how the application of our model may affect the interpretation of real 
astronomical data, we apply our treatment to the outer regions of the 114-426 
protoplanetary disk in Orion, previously analyzed by \cite{Miotello et al} in the 
framework of the standard LBB extinction law. 
The composition of the protoplanetary disk 
grains discussed above is identical to the one used by \cite{Miotello et al}.
It is therefore sufficient to adopt the same value of \cite{Miotello et al} for the intensity of the
illuminating source, assumed this time uniform over
a $2\pi$ hemisphere, to immediately derive and compare the values obtained 
for the  average grain size and optical depth. 

We refer in particular to the three pixels A, B, C, of \cite{Miotello et al},  located at the  
NE edge of the disk.  The results shown in Table~\ref{tab:Miotello} are relative to 
a) the average 
grain size; b) the product of the grain number density $n$ times the geometric 
thickness $L$ of the region, in units of $\mu$m$^{-2}$; this is the wavelength independent parameter directly returned by our model fitting, together with the grain radius 
$a$; and c) the adimensional optical thickness, 
given by the product $nL\sigma_g$. 

%consistently with the wavelength dependence of the cross section. 
Table~\ref{tab:Miotello} shows that the two methods return about the same average grain size, 
$a\simeq 0.5~\mu$m. 
%well within the uncertainties of our assumptiions. 
The other values, however, are significantly different. In particular the product $nL$,  that
basically represents the number of particles that have contributed to the observed 
extinction, increases by a factor $\simeq 1.3-2.9$, whereas the
optical thickness at $0.55~\mu$m
increases from $\tau\lesssim1$ to $\tau = 5.26, 3.28, 2.00$ for pixels A,B,C, respectively. 
Accounting for the extended background illumination makes the spectral energy distribution
emitted by the outer disk regions compatible with an optically thick medium. The column
density estimated using the LBB law may therefore represent a lower limit.

It must be remarked that in this specific case the outer disk regions may lie in the shadow of the thick disk,  i.e. the
illumination from the background may mostly come from less than $2\pi$~sr.
%%%%%
%Also, radiation can exit the disk borders because of the finite geometry. 
%%%%%
We have detected some indication of
this  "edge effect" in our experimental setup. Also, the emitting columns subtended by the 3 pixels
may not be well represented by the uniform plane-parallel slab we have assumed. Accounting for these effects would require a more
refined treatment of the geometry of the outer disk regions, which is still uncertain and beyond the scope
of this work.

\begin{deluxetable*}{lcccccc}[b!]
\tablecaption{Comparison of the results obtained with the classical LBB law by \cite{Miotello et al} and our model (indicated here as Radiative Transfer Model, RTM) for a) the the typical radius of the dust grains  (columns 1 and 2); b) the product $nL$ between the number density $n$ and the geometrical thickness of the 
scattering column $L$ (columns 3 and 4); c) the optical thickness.
\label{tab:Miotello}}
\tablecolumns{5}
\tablenum{2}
\tablewidth{0pt}
\tablehead{
\colhead{} &
\multicolumn{2}{c}{$a$[$\mu$m]} &
%\colhead{a[$\mu$m]} &
%\colhead{a[$\mu$m]} &
\multicolumn{2}{c}{$nL$[$\mu$m$^{-2}$]} &
%\colhead{nL[$\mu$m$^{-2}$]} &
%\colhead{nlL$\mu$m$^{-2}$]} &
\multicolumn{2}{c}{$\tau_{0.55}$} \\
%\colhead{$\tau_{0.55}$} &
%\colhead{$\tau_{0.55}$} \\
% & LBB & RTM & LBB & RTM & LBB & RTM  
 & \colhead{LBB} & \colhead{RTM} & \colhead{LBB} & \colhead{RTM} & \colhead{LBB} & \colhead{RTM}  
}
\startdata
Pixel A         & 0.6 & 0.6 & 0.59 & 1.71 & 0.98 & 5.28\\
Pixel B         & 0.5 & 0.5 & 0.55 & 1.45 & 0.95 & 3.28\\
Pixel C        & 0.4 & 0.5 & 0.83 & 1.12 & 0.73 & 2.00 \\
\enddata
%\tablenotetext{a}{At exposure start.}
\end{deluxetable*}

\section{Conclusions}

In this paper we have revisited the RTE theory for clouds of scatterers illuminated by an extended background source. We have derived a rigorous solution based on the assumption that multiple scattering produce an isotropic flux. We have then derived an approximate analytic model that nicely matches the results of the rigorous approach, and compared our predictions with accurate measurements for various types of well characterized scatterers, finding excellent match without the need of adding free parameters.

%Systematic effects due to the non ideality of the experimental setup have been recognized, related in particular to the asymptotic behavior for infinite optical thickness and the different angular aperture of the illuminating radiation.

We have used the predictions of our model to explore the behavior of an astrophysical cloud with dust grains having parameters, in terms of size distribution, composition, optical properties, typical of interstellar disks and the classic ISM grains of \cite{Draine}, illuminated by a diffuse source. We find that the cloud still exhibit properties of reddening, as expected on the basis of the LBB law, but they are appreciably modified in terms of transmission and shape of the spectral intensity distribution. In particular, for small optical thickness our results show that scattering makes reddening negligible at visible wavelengths. Once the optical thickness increases enough and the probability of scattering events become close to or larger than 1, reddening is appreciably modified by the amount of radiation coming from directions different that the line of sight and scattered towards the observer. Moreover, variations of the grain refractive index, in particular the amount of absorption, plays an important role also in changing the shape of the spectral transmission curve, with dielectric grain showing the minimum amount of reddening. The results presented in this work could also be useful for studying the atmospheres of extrasolar planets and brown dwarfs, where the presence of non--isotropic phase functions is the issue currently limiting analytic solutions of RTE \citep{bailey}.

\acknowledgements
The authors wish to acknowledge L. Ricci and W. Henney for early discussions on the radiative transfer through dark silhouette disks in the Orion Nebula. They are also indebted to anonymous referee for careful
review of the original manuscript and helpful comments.

\newpage 

\section*{Appendix}
Here we briefly present the methods we adopted to determine the quantities needed for the model to be compared to the results obtained with nonspherical particles, namely ceria oxide and carbon black.

We performed measurements with a commercial spectrophotometer (SP), a custom made laser turbidimeter (LT), a small angle laser light scattering (SALS) based upon the novel method of near field scattering (Mazzoni 2013), and a custom made optical particle counter (OPC). Thanks to these independent measurements, the scattering phase functions $p(\mu)$, the scattering optical depths $\tau_{sca}$, and the extinction optical depth $\tau_{ext}$ have been determined as a function of the wavelength as described below, thus completing the parameters needed to evaluate the extinction curve accordingly to our model.

The parameters have been determined as follows. The spectrophotometer provided the extinction of the sample as a function of the wavelength, $\tau_{sca}(\lambda)$. The same samples have also been measured with the LT for giving an absolute, precise check of the extinction efficiency measured with the SP at the wavelength of the laser ($\lambda_0$ = 632.8 nm). We adopted this additional check to be sure that the SP measurement was absolutely not affected by contributions coming from multiple scattering events. This contribution is rigorously get rid of with the LT, while in principle could be still present in the SP especially for the ceria oxide suspension (when the extinction is almost completely due to scattering with no absorption).

We then measured the samples with SALS covering a scattering wavevector range from 0.1 to 4 $\mu$m$^{-1}$. This range corresponds to an angular range wide enough to include almost the whole scattering lobe for both the samples, and thus to determine the phase functions $p(\mu)$. Moreover, the SALS measurements also provide an absolute measurement of the total scattering cross section, that immediately brings to $\tau_{sca}(\lambda_0)$ at the laser wavelength. In the case of the ceria oxide it results in accordance with the value measured with the SP (and the LT, which operates at a very similar wavelength of SALS), meaning that $\tau_{sca} (\lambda_0) = \tau_{ext} (\lambda_0)$ and $\tau_{abs} (\lambda_0) = 0$. By contrast, carbon black provided $\tau_{sca} (\lambda_0) < \tau_{ext} (\lambda_0)$, so that $\tau_{abs} (\lambda_0) > 0$ accordingly to the absorbing nature of the material.

Now we introduce a simple assumption for evaluating the spectral albedos: both ceria oxide and carbon black are endowed with uniform albedos overall the wavelength spectrum, accordingly to their white and black colors. As a result, it is now possible to estimate the $\tau_{sca} (\lambda)$ as follows:
\begin{equation}
\tau_{sca} (\lambda) = \tau_{ext} (\lambda)  \frac{\tau_{sca} (\lambda_0) }{\tau_{ext} (\lambda_0) }
\end{equation}
and therefore $\tau_{abs} (\lambda) = \tau_{ext} (\lambda) - \tau_{sca} (\lambda)$. In such a way the extinction curve provided by our model can be evaluated without any free parameter, as it has been shown in the text (see Fig. \ref{fig:figura8}).

\end{document}